\documentclass[a4paper,11pt]{article}

\usepackage[final]{graphicx}
\usepackage[font={small}, labelfont=bf, justification=centering]{caption}[2007/01/07]
\usepackage[numbers]{natbib}
\usepackage[utf8]{inputenc}
\usepackage{mathtools}
\usepackage{amsmath}
\usepackage{amsbsy}
\usepackage{caption}
\usepackage{amssymb}
\usepackage{bm}
\usepackage{authblk}
\usepackage{units}
\usepackage{cleveref}
\usepackage{color,soul}
\usepackage{float}
\usepackage{tikz}
\usepackage[titletoc]{appendix}

\DeclareMathAlphabet\mathbfcal{OMS}{cmsy}{b}{n}

\usepackage[final]{showlabels}

\usepackage{amssymb}  
\usepackage{amsthm}
\usepackage{natbib}
\usepackage{graphicx}
\usepackage{breqn}

\usepackage{booktabs}

\usepackage{placeins} 
\usepackage{subcaption}
\usepackage{framed}
\usepackage{enumerate}
\usepackage{cleveref}

\usepackage{soul}

\usepackage{scrextend}

\usepackage{mymacros_11Sep20}

\newcommand{\cvec}[1]{{\mathbf #1}}

\newcommand{\diff}[1]{\mbox{d}{#1}} 

\newcommand{\nfg}{N^{FG}} 
\newcommand{\ncg}{N^{CG}} 

\newcommand{\pot}{U} 
\newcommand{\Ham}{\mathcal{H}} 
\newcommand{\Veff}{V^{\mathrm{eff}}} 
\newcommand{\fprojbf}{{\mathbfcal{F}}} 

\newcommand{\expp}[1]{\mbox{exp}\sqp{#1}} %
\newcommand{\Zeta}{Z} 

\newcommand{\Aphase}[1]{\boldsymbol\zeta_{#1}} 
\newcommand{\Bphase}[1]{\boldsymbol\Zeta_{#1}} 

\newcommand{\Lio}{\mathfrak{L}} 
\newcommand{\pproj}{\mathcal{P}} 
\newcommand{\qproj}{\mathcal{Q}} 
\newcommand{\idproj}{\mathcal{I}} 

\renewcommand{\exp}[1]{\mathrm{e}^{#1}} 

\newcommand{\ts}{\tilde{s}} 
\newcommand{\ttt}{\tilde{t}}

\newcommand{\tBphase}{\Bphase{}^\prime}

\newcommand{\fricmat}{\mathcal{M}} 
\newcommand{\bfricmat}{\bs{\fricmat}} 

\usepackage[margin=0.8in]{geometry}
\usepackage{setspace}
\newgeometry{left=2.4cm,bottom=2.3cm,top=2.3cm,right=2.4cm}

\newtheorem{prop}{Proposition}
\crefname{prop}{proposition}{Proposition}
\Crefname{prop}{Propositions}{Propositions}

\graphicspath{ {Fig/} }

\begin{document}

\title{A Systematic Analysis of the Memory term in Coarse-Grained models: the case of the Markovian Approximation}
\author[1]{Nicodemo Di Pasquale \footnote{Authors listed in alphabetical order \\
\hspace*{0.4cm}$\dagger$ corresponding author: nicodemo.dipasquale@manchester.ac.uk}$^\dagger$}
\author[2]{Thomas Hudson}
\author[3]{Matteo Icardi}
\author[4]{Lorenzo Rovigatti}
\author[5]{Marco Spinaci}
\affil[1]{Department of Chemical Engineering and Analytical Science, University of Manchester, Manchester M13 9PL, UK}
\affil[2]{Warwick Mathematics Institute$,$ University of Warwick$,$ Coventry CV4 7AL$,$ UK}
\affil[3]{School of Mathematical Sciences$,$ University of Nottingham$,$ Nottingham NG7 2RD$,$ UK}
\affil[4]{Dipartimento di Fisica, Sapienza Università di Roma, Rome, 00185, Italy}
\affil[5]{Dipartimento di Scienze Molecolari e Nanosistemi, Università Ca' Foscari, 30123 Venezia, Italy}

\date{}
\maketitle

\begin{abstract}
  The systematic development of Coarse-Grained (CG) models via the Mori-Zwanzig projector operator formalism requires the explicit description of a deterministic drift term, a dissipative memory term and a random fluctuation term. The memory and fluctuating terms are related by the fluctuation--dissipation relation and are more challenging to sample and describe than the drift term due to complex dependence on space and time.
  This work proposes a rational basis for a Markovian data-driven approach to approximating the memory and fluctuating terms, and its use is demonstrated through an application to the estimation of the memory in the coarse-grained model of a one-dimensional Lennard-Jones chain with different masses and interactions.
\end{abstract}

\section{Introduction}
The problem of predicting the evolution of a non-linear dynamical system is ubiquitous in science and engineering. In many such applications, the fact that thousands or millions degrees of freedom must be solved for at once makes direct numerical solution infeasible, and so many methods are constrained to apply to smaller (and hence less scientifically interesting) problems.
One of the most relevant example of this kind is represented by Molecular Dynamics (MD) simulation \citep{leimkuhler2016molecular}. In an MD simulation a set of atoms which represents a particular molecular system is considered, and their positions and momenta are calculated, assuming that they evolve according to the Newton's equations of motion~\cite{rapaport2004art} with prescribed interaction potentials and an appropriate thermostat. 

Fully atomistic MD simulation regularly performed nowadays reach the order of a few million atoms ($\approx 10^6$) \citep{Zhao2013}, a number which remains well below the size of a macroscopic system with Avogadro's number of atoms ($\approx 10^{23}$). Although simulations with around $10^4$ atoms are usually large enough to extract properties of interest in systems like crystals and small macromolecules (such as polymers and peptides), to understand the dynamics of proteins, one must consider the environment in which they evolve. From the point of view of simulation, the problem for this kind of systems is not only the number of degrees of freedom (DoFs) required to represent the system, but also the relaxation times of these bio-macromolecules. These can be so large that the cost of performing any atomistic simulation becomes quickly prohibitive, since billions of time steps are still required to extract meaningful information. This dual space-time problem continues to render accurate simulations of this type out of our reach.

For these reasons, in recent years a series of techniques to derive simpler Coarse-Grained (CG) model have been developed. The main idea behind CG models is to reduce the number of variables in the system while maintaining accuracy. In place of the detailed evolution of the atoms, we consider a smaller number of macroscopic variables, functions of the underlying atomistic structure~\cite{voth2008coarse,papoian2017coarse}. Although theoretical results concerning static equilibrium statistics for coarse-grained systems are relatively well-developed \citep{lelievre2010}, theoretical work on dynamical CG models is still evolving, particularly focusing on applications of the Mori-Zwanzig (MZ) projection operator formalism. In \citep{Hijon2010}, a systematic derivation of CG models, based on MZ techniques was proposed. The authors of the present work contributed by putting various aspects of MZ-CG models on a more rigorous footing \citep{DiPasquale2019}. The idea behind applying the MZ framework to coarse-graining is to follow the dynamics of the CG DoF by explicitly integrating over the non-coarse-grained DoF. The resulting equations represent a formidable mathematical problem, and several works were already published reporting the analysis of different aspect of the MZ equations  \citep{Givon2005,Chorin2009,Chorin1998}. The importance of the MZ equations is not limited to CG problems alone, and similar approaches have been proposed to describe many different systems, ranging from the relativistic heavy-ion collision, where Fick's law breaks down \citep{Koide2005,Koide2007}, to the dynamics of supercooled liquids, glasses and other amorphous solids~\cite{zaccarelli2001gaussian,voigtmann2004tagged,gotze2008complex,Cui2017,Cui2017b,Cui2018,Cui2020,handle2019q,Zaccone2020}.


\subsection{The Mori-Zwanzig equations}
The MZ-CG equations are a set of integrodifferential equations which take the form of Generalized Langevin Equations under appropriate assumptions.
In a CG model, these equations for $\ncg$ beads can be written as (for a full derivations see \citep{DiPasquale2019,Hijon2010}):
\begin{align}
 \nonumber
 \totd{\Bphase{\bfR}}{t} &= \cvec{M}^{-1}\Bphase{\bfP}\,; \\
 \totd{\Bphase{\bfP}}{t} &= - \pard{\Veff}{\Bphase{\bfR}} 
    + \int_{0}^{t}  \exp{(t-s)\,\Lio}\,\pproj\Lio\,\exp{s\,\qproj\Lio}\,\qproj\Lio\,\Bphase{}\,\de s 
    + \fprojbf_{\Bphase{\bfP}}(t,\cdot)\,;
 \label{Eq:GLE_CGs} 
\end{align}
where:
\begin{itemize}
    \item $\Bphase{}=\left(\Bphase{\bfR},\Bphase{\bfP}\right)=\left(\bfR_1,\ldots,\bfR_\ncg,\bfP_1,\ldots,\bfP_\ncg\right)$ are the CG DoF;
    \item $\cvec{M}=\mathrm{diag}(M_1,\dots,M_{\ncg})$ is a diagonal matrix of the masses of the beads;
    \item $\Veff$ is the effective potential (for an explicit definition we refer to \citep{DiPasquale2019});
    \item $\Lio$ is the Liouvillean operator; and
    \item $\pproj$ and $\qproj$ are the MZ projection operator and its orthogonal counterpart (i.e. $\qproj=\idproj-\pproj$) respectively.
\end{itemize}   Finally, the term $\fprojbf_{\Bphase{\bfP}}(t,\cdot)$ is often called \emph{fluctuating force} and can be written (for the $K$-th bead) as
\begin{equation}\label{Eq:FluctuatingForce}
	 \fprojbf_K(t,\cdot) =  \,\exp{t\,\qproj\Lio}\,\qproj\Lio\,\cvec{P}_K \,. 
\end{equation}
Here, we used a dot in the second argument of $\fprojbf_K$ to stress that this term is a function of all of the variables in the atomistic system.
Computing $\fprojbf_K$ requires the calculation of the evolution of the orthogonal variables in the null space of $\pproj$. In general, it is given by solving an auxiliary set of equations called \emph{orthogonal dynamics equations} \citep{Givon2005}, which (with an appropriate CG mapping) can be determined by means of constrained dynamics (see \citep{DiPasquale2019}). Its presence in the MZ equations \cref{Eq:GLE_CGs} is usually considered as a random noise for the CG DoF \citep{Chorin2000}. Making this assumption avoids the problem of its direct calculation, and so allows us to reduce the complexity of the system. However, ensuring that statistics of $\fprojbf_K$ are captured accurately in a stochastic model is important to preserve the accuracy of the CG system. This represents the main problem we address in this work.

The integral term in \cref{Eq:GLE_CGs}  is known as the ``memory'', since requires information coming from the elapsed time. In a sense, in order to know the next position occupied by the system we need to ``remember'' what happened since the initial time $t=0$.
The memory term can be rewritten (for the $K$-th bead) as \citep{DiPasquale2019,Hijon2010}
\begin{align}\label{Eq:MemTerm}
	 \int_{0}^{t}  \left[\exp{(t-s)\Lio}\,\pproj\Lio\,\exp{s\,\qproj\Lio}\,\qproj\Lio\,\Bphase{}\right]_K\,\de s = \sum_{J=1}^{\ncg} \int_{0}^{t} \fricmat_{KJ}(\Bphase{}(t-s),s) \,\frac{\cvec{P}_J(t-s)}{M_J} \,\diff{s}  \nonumber \\
    + \beta^{-1} \sum_{J=1}^{\ncg} \int_{0}^{t} \pard{\fricmat_{KJ}(\Bphase{}(t-s),s)}{\cvec{P}_{J}} \,\diff{s} 
\end{align}
where we introduced the $KJ$-th components of the memory kernel $ \bs{\fricmat}\cip{\Bphase{}(t-s),s} $:
\begin{align} \label{Eq:fricmat}
 \fricmat_{KJ}\cip{\Bphase{}(t-s),s} =& \,\beta\,\pproj\sqp{ \cip{ \qproj\Lio\,\cvec{P}_K } \otimes \cip{ \exp{s\,\qproj\Lio}\,\qproj\Lio\,\cvec{P}_J } }  {}  \nonumber\\
 =& \,\beta\, \pproj \sqp{ \fprojbf_K(0,\cdot) \otimes \fprojbf_J(s,\cdot) } \nonumber \\
 \equiv& \,\beta\, \cexpval{ \fprojbf_K(0,\cdot) \otimes \fprojbf_J(s,\cdot) }{ \,\Bphase{}(t-s) } \,. 
\end{align}
Here, $\otimes$ represents a tensor product and $\beta^{-1}=\kb T$, with $\kb$ the Boltzmann constant, and
$\fprojbf_K(0,\cdot) = \qproj\Lio\,\cvec{P}_K\,$ (see \cref{Eq:FluctuatingForce}).
Di Pasquale \textit{et al.} have shown that (with an appropriate CG mapping) the latter term is given by the difference of the total force acting on atoms within a bead and the mean force given by the effective potential that describe the bead interaction (see Ref.~\citep{DiPasquale2019}). Finally in the last passage, we used the Zwanzig projection $\pproj$, and the fact that it is equivalent to take the conditional expectation given the CG (set of) variables $\Bphase{}$, interpreted as random variables, with respect to the equilibrium distribution~\citep{Chorin2000,DiPasquale2019}.

For a system in thermal equilibrium, the memory kernel is related to the autocovariance of the fluctuating forces (ACF) through the \textit{Fluctuation-dissipation theorem} \cite{Jung2017,Farias2009,Hanggi1995}:
\begin{equation}\label{Eq:flucDiss}
\beta\aver{\fprojbf(0)\otimes\fprojbf(t)} = \bfricmat(t) \,, 
\end{equation}
where $\aver{\cdot}$ denotes an ensemble average.
This description is usually called ``colored noise'' fluctuation, as opposed to the ``white noise'' represented by the Markovian behaviour (see \cref{Sec:Markovian}) \citep{Hanggi1995}. 

Different approaches have been proposed in the literature to deal with the memory term. In ~\citep{Kauzlaric2011}, an analysis of the memory term obtained from the Mori projector operator \citep{Mori1965,Kawasaki1973}, a special case of the more general MZ projector, was presented. A more detailed analysis is reported in \citep{Zhu2018} were some boundary analysis is proposed and the hierarchical approximation, first introduced by \citep{Stinis2007}, is further developed. Other approaches proposed in the literature to calculate the memory term, include the use of the Galerkin decomposition \citep{Darve2009}, or the assumption that the evolution of the CG DoF can be modelled using an autoregressive stochastic process \citep{Kneller2001}.

The two most used approximations for the calculation of the memory term are related to the behaviour of the integrand of the memory integral with respect to time. If the integrand quickly decays  to zero, it can be assumed that all the information is concentrated in a very short time (i.e. we don't need the whole history of the system from $t=0$). The memory term becomes local in time and we talk about the \textit{Markovian Approximation} \citep{Hijon2010,Chorin2009}.
If, on the other hand, the decay of the integrand of the memory term is very slow compared to the characteristic time of the dynamics of the system, then we can assume that the integrand is constant in the interval $[0,t]$ and we obtain the so-called t-model \citep{Chorin2002,Hald2007,Chorin2009,Price2019}.

In this work we will focus on the short-time approximation by describing a model which can be interpreted as giving a Markovian reduction of the MZ equation.
This paper is structured as follows: we will start with the discussion of the memory term from the point of view of the Markovian Approximation as usually is considered in the literature, and then we present our new description for it. We will show how it can be effectively approximated and we will give a precise definition of all the hypothesis used in the derivation. We will then apply the framework derived in a simple example composed by a one dimensional linear chain where atoms interact with a non-linear potential. We will show how the relevant quantities described here can be explicitly calculated and we draw some conclusions.

\section{The behaviour of the MZ CG system}

\noindent A significant challenge with handling the memory term arises due to the need to approximate the memory kernel $\bfricmat$, which varies in both time and space. A route often followed in the literature is to assume a functional form for either the whole memory kernel or just the fluctuating force, and to ensure that both are coupled through the Fluctuation-Dissipation Theorem. In \citet{Berkowitz1983} the starting point is to assume that the fluctuating force is a periodic function which can thus then developed in Fourier series. Taking the other perspective, functional forms for the memory kernel that have been considered include Gauss-exponential process \citep{Berkowitz1983,Berne1970,Wang1945},  the Gauss-Gauss process \citep{Berkowitz1983}, or combination of exponentially decaying kernels \citep{Berkowitz1983,Ayaz2021}.
In all these cited examples, however, the choice of ansatz implies a strong hypothesis: the memory kernel does not depend on the positions (i.e. it is spatial homogeneous). However, as we will show in the next sections, for most CG systems this hypothesis does not seem justified. Even in a 1-dimensional chain interacting with a simple Lennard-Jones potential, which is the example we present here, we observe that the memory kernel has a strong spatial dependence. We may therefore infer that for a more complicated system in higher dimensions, this spatial dependence will be even more important to consider. 

\subsection{Memory approximation}\label{Sec:Markovian}

The physical theory of Brownian motion assumes that thermal fluctuations which occur due to the collision of the small, light particles with the larger Brownian particle happen at a time-scale which is much shorter than the one describing the evolution of the larger particle. The consequence of this assumption is that the force acting on the Brownian particle at each time is completely uncorrelated with the history of its evolution. Mathematically-speaking, this assumption corresponds to the idea that the random force is a stochastic process which obeys the Markov property. We note that in the theory of Brownian motion, there is a further important assumption, which is that the behaviour of the heat bath is spatially uniform. This assumption is clearly justified for a particle within a fluid which is at rest macroscopically.

For coarse-graining of molecular systems, it is common to make an analogous time-scale separation assumption about the action of the neglected DoF on the coarse-grained DoF, i.e. that the discarded DoF fluctuate at a faster time-scale and so act on the CG DoF randomly with no correlations in time. On the other hand, the latter assumption of spatial uniformity is however not generally valid, since many coarse-grained DoFs in this context are tightly bound within a molecule or bulk solid, and so are in close contact with their particular environment, rather than a homogeneous random one.

Setting aside this issue for the moment, under the assumption of time-scale separation and spatial homogeneity, the fluctuating forces $\fprojbf$ are usually modelled as a white noise process, with an ACF which is assumed to take the form:
\begin{equation}\label{Eq:statPhi}
	\beta\,\mathbb{E}[\fprojbf_K(0)\otimes\fprojbf_J(t)] = 2\,\Phi_{KJ} \dirac{t}\quad\text{where}\quad \Phi_{KJ} = \beta\int_{0}^{\infty}  \mathbb{E}[\widetilde{\fprojbf}_K(0)\otimes\widetilde{\fprojbf}_J(\xi)] \de \xi,
\end{equation}
with $\widetilde{\fprojbf}$ being the fluctuating forces which act in the system at full resolution.
Since the fluctuating forces are assumed to be stationary as a random process (in fact they are defined as averages over equilibrium ensembles, see \citep{Berne:DynamicLightScatt}), the terms $\Phi_{KJ}$ do not depend on the particular time origin \citep{Lei2010,Kinjo2007}, i.e. there is no time dependence in the definition of $\Phi_{KJ}$ given in \eqref{Eq:statPhi}.
%
%
One of the problems with this description is the so-called \textit{plateau problem}, first noticed by \citet{Kirkwood1949}. The plateau problem stems from the fact that the integral definition of $\Phi_{KJ}$ is zero if the upper boundary is $+\infty$, because of the decay of the negative part of the correlation function at infinity \citep{Helfand1961}. For this reason, one approach has been to cut the domain of integration for the last integral in \cref{Eq:statPhi} at a certain time $\tau$ which should be large enough to include the relevant information in the ACF, but small compared to the time scale of the evolution of the CG DoF. By cutting the integral at a finite time we are assuming that it remains constant for $t > \tau$ (i.e. the integral reaches a plateau). The problem comes from the fact that the intermediate time $\tau$ is not well defined within the theory and there is no method to estimate it \textit{a priori}.

In the following we will show that our derivation overcomes this issue by proposing an asymptotic methodology which allows us to use data to fit a functional form for the memory, and thereby derive an explicit form for the Markovian approximation directly from data.

\subsection{Memory ansatz}\label{sec:analysisMemory}

Our analysis starts in a similar fashion by making an ansatz on the form of the memory kernel $ \bs{\fricmat}(\Bphase{},t)$, albeit a less restrictive one. We assume that the memory kernel for the generic pair of beads $K$ and $J$ takes the general form
\begin{equation}
  \fricmat_{KJ}(\Bphase{},t) = \mbox{exp}\sqp{-\cip{\frac{t}{\tau_{KJ}}}^\alpha}f_{KJ}(\Bphase{},t),
\end{equation}
where $\tau_{KJ}$ is a characteristic time, $\alpha$ is a constant, and $f_{KJ}(\Bphase{},t)$ is a sufficiently smooth function on the interval $0\leq t < \infty$.  
In general, each entry of the memory kernel is specific to each pair of beads. However, in the following, in order to simplify the notation, we will write
\begin{equation}\label{Eq:ansatz}
  \bs{\fricmat}(\Bphase{},t) = \exp{- \cip{\frac  t\tau}^\alpha}f(\Bphase{},t),
\end{equation}
i.e. we will consider the case where all the entries in the memory kernel are equal. The general case can then be easily obtained by repeating the same arguments for all the possible $KJ$ pairs. It is interesting to note that a similar ansatz was considered when using a MZ-derived Generalized Langevin Equation to study the dynamical behaviour of amorphous solids \citep{Cui2017,Cui2017b,Cui2018,Cui2020,Zaccone2020}.

While \cref{Eq:ansatz} represents a somewhat strict ansatz for the form of the memory kernel, it is consistent with similar quantities calculated in other systems \citep{Li2017}, and moreover, our approach includes a possibly non-uniform dependence on the spatial position, in contrast to assumptions made in other applications of CG theory discussed above.
Moreover, given the connection between the memory kernel and the ACF of \cref{Eq:flucDiss}, both $\tau$ and $\alpha$ can be extracted from the ACF, for example by fitting to numerical data. In particular, $\tau$ represents the time decay of the ACF and we will show in the \cref{Sec:Res} its explicit calculation in a test system. We will also make the natural physical assumptions that $\bs{\fricmat}(\Bphase{},t)$ is maximal when $t=0$, and that the parameters satisfy $\tau>0$ and $\alpha > 0$.

Under the assumptions that $\alpha>0$ and $\tau>0$, the exponential factor makes it possible to expand the memory kernel as an asymptotic series from which the behaviour close to the origin can be explicitly obtained.
Here, we argue that this asymptotic series approach leads naturally to a form of the Markovian approximation commonly considered in CG theory. We note that the case where $\alpha = 0$ corresponds to algebraic decay of the memory kernel, and will require a different approach we leave for future work.


In the following, we consider only the first integral of the memory term (\textit{i.e.}, \cref{Eq:MemTerm}) since $\bfricmat$ does not depend on the momenta for an appropriate CG mapping, see \citep{DiPasquale2019}. Furthermore, we make the memory term dimensionless by multiplying it by suitable constants:
\begin{align}\label{eq:adim}
  &\frac{M L_c }{\tau_P^2}\int_0^{\tilde{t}}\, \bs{\widetilde{\fricmat}}\big(\tBphase{}(\tilde{t}-\tilde{s}),\tilde{s}\big) \cdot \widetilde{\bfP}(\tilde{t}-\tilde{s}) \,\diff{\tilde{s}} \\
  &\qquad = \int_0^{\tilde{t}} \, \exp{-\cip{\frac{\tau_P}{\tau}}^\alpha \tilde{s}^\alpha} \tilde{f}\big(\tBphase{}(\tilde{t}-\tilde{s}),\tilde{s}\big)  \widetilde{\bfP}(\tilde{t}-\tilde{s}) \,\diff{\tilde{s}} \nonumber
\end{align}
where:
\begin{itemize}
    \item $M$ is the mean effective mass of the coarse-grained degrees of freedom;
    \item $L_c$ is a characteristic length whose specific definition is not important for what is discussed here (and can be taken as the average bead size, for instance);
    \item $\tau_P$ is a characteristic time, which can be interpreted as the decay of the autocovariance function of the momenta (ACM) of the beads. This quantity can be interpreted also as the characteristic time-scale of the macroscopic system.
\end{itemize}
  We indicate with $\widetilde{\cdot}$ the dimensionless quantities and we note that we can always rewrite the different terms in \cref{eq:adim} as function of the dimensionless time $\ttt = t/\tau_P$, by writing $\Bphase{}(t-s)=\Bphase{}(\tau_P(\ttt-\ts))=\tBphase{}(\ttt-\ts)$, where we absorb $\tau_P$ in the definition of the function.
We will also simplify the notation in \cref{eq:adim} by defining the dimensionless quantity $\lambda=\cip{\frac{\tau_P}{\tau}}^\alpha$. We want to highlight here the first result of the formulation presented here. The concept of the time-scale separation can be precisely and quantitatively defined by measuring it with the parameter $\lambda$. From the definition of $\lambda$ we can distinguish two cases:
\begin{itemize}
    \item[$i)$]$\lambda \gg 1$, then $\tau_P \gg \tau$, that is to say that the macroscopic time-scale is much larger than the microscopic one. The Markovian approximation is expected to hold.
    \item[$ii)$] $\lambda \approx 1$, then $\tau_P \approx \tau$ and there is no separation between the macroscopic and the microscopic time-scale. The Markovian approximation  fails and the full memory term must be considered.
\end{itemize}

In the next section we will present a result which will make more clear the qualitative explanation given here in terms of validity or not of the Markovian approximation. Then, the problem will become the explicit calculation of the two quantities, $\tau$ and $\tau_P$, which will be presented in the following sections. In order to simplify our notation, we will drop tildes from now on, and assume all quantities are dimensionless.

\subsection{Rigorous approximation result} \label{Sec:alphaPos}


Under the assumption that the ansatz \cref{Eq:ansatz} holds, we now derive an error estimate for an expansion of the memory integral which we can use to define proxy dynamics, providing a theoretical guarantee of accuracy in the case where the time-scale separation is significant.

\begin{prop}\label{prop}
Suppose that the memory kernel $\fricmat$ takes the form
\[
\fricmat(\Bphase{}(t-s),s) = \exp{- \lambda s^\alpha}f(\Bphase{}(t-s),s),
\]
where $\lambda>0$, $\alpha>0$, and the function
\[
g(t,s) = f(\Bphase{}(t-s),s)\cvec{P}(t-s)
\]
is of class $\mathrm{C}^{k,\beta}$ in both $t$ and $s$, for some non-negative integer $k$ and H\"older exponent $\beta\in(0,1]$. Then the absolute error committed by replacing the memory 
\[
\int_0^t \fricmat(\Bphase{}(t-s),s)\cvec{P}(t-s) \diff{s} = \int_0^t \exp{- \lambda s^\alpha} g(t,s) \diff{s}
\]by a $k$-term approximation is bounded by
\[
  \Bigg|\int_0^t \exp{- \lambda s^\alpha} g(t,s)\diff{s} - \sum_{n=0}^{k} \frac{G_n(t)}{n!}\frac{\Gamma\big(\frac{n+1}{\alpha}\big)}{\alpha \lambda^{\frac{n+1}{\alpha}}}\Bigg|
  \leq \sum_{n=0}^{k}\frac{C_n}{n!}\int_t^\infty \exp{-\lambda s^\alpha} s^n \diff{s}+C_{k+1}\frac{\Gamma\Big(\frac{k+\beta+1}{\alpha}\Big)}{k!\alpha \lambda^{\frac{k+\beta+1}{\alpha}}},
\]
where
\begin{equation*}
G_i(t) = \frac{\partial^i}{\partial s^i}g(t,0)=\frac{\partial^i}{\partial s^i}\Big(f(\Bphase{}(t-s),s)\cvec{P}(t-s)\Big)\bigg|_{s=0},
\end{equation*}
$\Gamma$ is the Gamma function, and $C_n$ are appropriate positive constants.
\end{prop}
A full proof of this proposition is given in Appendix~\ref{proof}, and we focus on the physical interpretation of the result here.
The error bound for the memory matrix is composed of two terms. The first of these is
\begin{equation}
    \psi(t,\lambda) \overset{\operatorname{def}}{=} \sum_{n=0}^{k}\frac{C_n}{n!}\int_t^\infty \exp{-\lambda s^\alpha} s^n \diff{s},\label{eq:psi}
\end{equation}
and it can be shown that $\psi$ decays exponentially in time for fixed $\lambda$ (see \cref{app:psi}).
Physically, this decay represent transient behaviour as the system relaxes towards the equilibrium. We believe this transient behaviour is closely related to the plateau problem which has been observed in earlier works referred to above.

The second term,
\begin{equation}\label{eq:zeta}
    \zeta(\lambda) \overset{\operatorname{def}}{=} C_{k+1}\frac{\Gamma\Big(\frac{k+\beta+1}{\alpha}\Big)}{k!\alpha \lambda^{\frac{k+\beta+1}{\alpha}}}
\end{equation}
does not decay with increasing time $t$. However, we can study its behaviour with respect to $\lambda$. In particular, it follows that $\zeta(\lambda)$ decays to zero as $\lambda\to\infty$ because $\frac{k+\beta+1}{\alpha} > 0$, then 
\begin{equation}\label{Eq:zetaO}
\zeta(\lambda) = \mathcal{O}\cip{\lambda^{-\frac{k+\beta+1}{\alpha}}} \,, \quad \, \lambda\to\infty.
\end{equation}
As already observed (see \cref{sec:analysisMemory}), $\lambda$ represents the scale-separation between the underlying atomistic model and the coarse-grained one, so as we will show below, the above result provides both a form for a Markovian approximation of the dynamics, and a theoretical guarantee backing up the commonly-held belief that a Markovian approximation is more accurate when there is a large timescale-separation between atomistic and CG models.

In particular, as a corollary of \cref{prop}, we note that if the integrand function $g(t,s)$ in the memory integral is Lipschitz, so that $k=0$ and $\beta=1$, then we have
\[
\Bigg|\int_0^t \exp{- \lambda s^\alpha} g(t,s)\diff{s} - G_0(t)\frac{\Gamma\big(\frac{1}{\alpha}\big)}{\alpha \lambda^{\frac{1}{\alpha}}}\Bigg|\leq C_0\int_t^\infty \exp{-\lambda s^\alpha}\diff{s}+C_1\frac{\Gamma\big(\frac2\alpha\big)}{\alpha \lambda^{\frac2\alpha}}.
\]
We note that the function $G_0(t)$ is
\begin{equation*}
    G_0(t) = f(\Bphase{}(t),0)\cvec{P}(t) = \beta \cexpval{ \fprojbf(0,\cdot) \otimes \fprojbf(0,\cdot) }{ \,\Bphase{}(t) }\cvec{P}(t),
\end{equation*}
and hence we are able to approximate \cref{eq:adim}, where we have dropped the tildes, as
\begin{align}\label{Eq:approxM}
    \int_0^t \fricmat(\Bphase{}(t-s),s)\cvec{P}(t-s) \diff{s} & \approx  \frac{\Gamma\big(\frac{1}{\alpha}\big)}{\alpha \lambda^{\frac{1}{\alpha}}} \beta\, \cexpval{ \fprojbf(0,\cdot) \otimes \fprojbf(0,\cdot) }{ \,\Bphase{}(t) }\cvec{P}(t),
    \nonumber \\
    & \equiv \bs\chi(\Bphase{}(t);\alpha,\lambda )\cvec{P}(t).
\end{align}
or in other words, we can define a spatially-varying friction cooefficient matrix $\bs\chi$ to replace the memory integral.

In this case, we can use this expression to show that the relative error of the approximation is bounded by
\begin{equation}\label{eq:relative}
    \Bigg|\frac{\int_0^t \exp{- \lambda s^\alpha} g(t,s)\diff{s} - G_0(t)\Gamma\big(\frac{1}{\alpha}\big)\alpha^{-1} \lambda^{-\frac{1}{\alpha}}}{G_0(t)\Gamma\big(\frac{1}{\alpha}\big)\alpha^{-1} \lambda^{-\frac{1}{\alpha}}}\Bigg|
    \leq \frac{\int_t^\infty \exp{-\lambda s^\alpha}\diff{s}}{\int_0^\infty \exp{-\lambda s^\alpha} \diff{s}}+\frac{C_1}{\lambda^{\frac1\alpha}|G_0(t)|}\frac{\Gamma\big(\frac2\alpha\big)}{\Gamma(\frac1\alpha)}.
\end{equation}
As we already observed for the absolute error bound, the former term exhibits decays to zero as $t\to\infty$, regardless of the value of $\lambda$, while the latter decays as $\lambda\to\infty$ for fixed $t$.


In the next section we will show how to compute $\chi$ in practice, and discuss the error committed in replacing the memory term with its approximation shown in \cref{Eq:approxM}.

\section{A numerical example}
In this section, we demonstrate how one might apply the mathematical result of Proposition~\ref{prop}. To do so, we use a numerical implementation of a one-dimensional periodic chain made of $\nfg$ atoms of two different species, which differ in mass and bond stiffness.

\subsection{Model}
We assume the atoms are arranged in a repeating pattern, and a coarse-grained model is obtained by combining the single repeated units into beads. The CG variables are given by the centre of mass of each bead and by the corresponding momenta, and the repeating pattern consists of two atoms of mass $M_2$ and a single atom of mass $M_1$ arranged in the following configuration: $(M_2-M_1-M_2)$. We considered two test cases: $M_1=M_2=1$ and $M_1=1$, $M_2=100$.

The bond stiffness between the two different species of atoms is described by means of an inter-atomic potential, which is a simple 12-6 Lennard-Jones potential:
\[
  	\pot_{i,i+1}(r) = 4\,\epsilon_{i,i+1} \sqp{\cip{\frac{\sigma_{i,i+1}}{r}}^{12} - \cip{\frac{\sigma_{i,i+1}}{r}}^6} = \epsilon_{i,i+1} \sqp{\cip{\frac{r^*_{i,i+1}}{r}}^{12} - 2\cip{\frac{r^*_{i,i+1}}{r}}^6} \,.
\] 
Here, $i \in \bp{1,2,\dots,\nfg}$ and we choose $\sigma_{i,i+1}=2^{-1/6}$ so that the minimum value of the potential is obtained for $r^*_{i,i+1}=1$. The value of the interaction strength, $\epsilon_{i,i+1}$, is set to $1$ and $10$ for $M_1 - M_1$ and $M_1 - M_2$ pairs, respectively. 
In the following, all parameters, such as temperature and time step, are expressed in terms of Lennard-Jones reduced units \citep{tuckerman2010}.

The simulations to obtain the fluctuating force (and therefore to determine the memory kernel) are implemented in the canonical (NVT) ensemble at fixed $k_B T = 1$ and friction parameter $\gamma=1$. Furthermore, we assume periodic boundary conditions and nearest neighbour interactions. 

The Hamiltonian of the atomistic system is given by
\begin{equation}
 \Ham(\cvec{r},\cvec{p}) = \frac12 \,\cvec{p}^T\,\cvec{M}^{-1}\,\cvec{p}\,+\sum_{i=1}^{\nfg-1}\pot_{i,i+1}\cip{r_{i+1}-r_i}\,+ \pot_{\nfg,1}\cip{r_1-r_{\nfg}+r^*\nfg},
 \label{eq:1DHamiltonian}
\end{equation}
where $\cvec{M}$ is the mass matrix (that is, a diagonal matrix whose entries are the masses of each atom) and $\cvec{r}$ and $\cvec{p}$ are the positions and momenta of the atoms, respectively. 
The final term takes into account that periodic boundary conditions are in place, \textit{i.e.}, that the two ends of the linear chain are connected to each other.

The results shown below were obtained through simulation of two different dynamics: (i) the  \emph{Fine-Grained Dynamics} (FGD) when solving the equations of motion with the Hamiltonian defined in eq.~(\ref{eq:1DHamiltonian}),
 \[
 \totd{\Aphase{}(t)}{t} =  \Lio\,\Aphase{} \,,
 \]
 where, $\Aphase{}=\cip{ \Aphase{ \cvec{r} },\Aphase{ \cvec{p} } }=( \cvec{r}_1,\ldots,\cvec{r}_{\nfg} , \cvec{p}_1,\ldots,\cvec{p}_{\nfg} )$ is a state of the system characterized by the instantaneous positions and momenta of the $\nfg$ particles that compose it; (ii)  the \emph{Orthogonal Dynamics} (OD) when solving the system of equations
 \[
  \totd{r_k}{t} = \frac{p_k}{m_{k}}- \frac{P_I}{M_I} \,; \qquad
  \totd{p_k}{t} = - \pard{\pot(\Aphase{\cvec{r}})}{r_k} + \frac{m_k}{M_I}\sum_{i\in S_I}\pard{\pot(\Aphase{\cvec{r}})}{r_i}\,;
 \]
 for $k = 1,2,\dots,\nfg$, where $I$ is the index such that $k \in S_I$ and $S_I$ is the set of particles of the FG system included within the bead of index $I$.

\subsection{Numerical sampling of the memory kernel}

All the simulations reported below were performed using the code \emph{HybridZwanzJulia}  \citep{MZcode} written in Julia v1.4.1 \citep{Bezanson2017}. This code allows a relatively simple recalibration of the test system through Julia “types”, which are initialized by the user to fix all the parameters of the simulation, including the repeating pattern of the beads, the number of beads, the mass and stiffness of the inter-atomic potentials. Here, we give an idea of the sampling implemented in the code to compute the fluctuating force, given in \cref{Eq:FluctuatingForce}, and therefore the memory term as given in \cref{Eq:approxM}.

As mentioned in the introduction, the idea is to sample using OD, since it is possible to determine the fluctuating force through constrained dynamics.
First, the initial conditions for the simulation are fixed, so that the momenta are distributed according to the Maxwell-Boltzmann distribution and the positions lie in a neighbourhood of the equilibrium positions, determined by the equilibrium distance of the inter-atomic potential. 
Then the trajectory of the FGD is computed and samples of the positions and momenta are stored. These samples are used to determine the CG positions and momenta, and are used as initial condition for computing the trajectories of the OD. 
Along these trajectories, time averages of the first and second moment of the effective force between beads are stored. 
The first moment gives the value of the mean force acting between adjacent beads given the position of the center of mass, while the second moment is used to compute the fluctuating force. 
Since the sampling steps of the OD can run at the same time without having to communicate between the processes, these procedures can be run in parallel.

In order to improve the time required to sample the fluctuating force, we use the translational symmetry of this simple system, the fact that the beads are identical and the assumption that the effective potential can be approximated as a sum of two-body interactions based on the distance to the nearest-neighbour beads only. 
We thus compute the interactions between any combination of beads and assume that the interactions computed are valid for any equivalent pair in the system. 
In this way we reduce the dimensionality of the space to be sampled from the total number of coarse-grained variables to simply one, \textit{i.e.}, the inter-bead distance, which simplifies the exploration of the entire constrained phase space.

If we consider bead $I$, we can assume that the forces that act on it can be decomposed as the independent ``stress'' contributions $\sigma_{I+1,I}$ and $-\sigma_{I,I-1}$, which are the forces exerted on bead $I$ by its two neighbours, $I-1$ and $I+1$. 
Under these conditions we write the conditional expectation in the \cref{Eq:approxM} as
\begin{align*}
  \cexpval{\fprojbf_I \otimes \fprojbf_I}{\cvec{R}} &= 
  \cexpval{ (\Delta\sigma_{I+1,I}-\Delta\sigma_{I,I-1}) \otimes (\Delta\sigma_{I+1,I}-\Delta\sigma_{I,I-1}) }{\cvec{R}} = \\
  &= \cexpval{ \Delta\sigma_{I+1,I} \otimes \Delta\sigma_{I+1,I} }{\cvec{R}} + \cexpval{ \Delta\sigma_{I,I-1} \otimes \Delta\sigma_{I,I-1} }{\cvec{R}} \,,
\end{align*}
where $\Delta\sigma$ is the fluctuating part of the ``stress'' contribution, $\cvec{R}$ is the inter-bead distance, and we assume that the cross terms are negligible, \textit{i.e.},
\[
 \cexpval{ \Delta\sigma_{I+1,I} \otimes \Delta\sigma_{I,I-1} }{\cvec{R}}\approx 0 \quad\text{and}\quad \cexpval{ \Delta\sigma_{I,I-1} \otimes \Delta\sigma_{I+1,I}}{\cvec{R}}\approx 0\,.
\]
In other words, we assume that the fluctuating stresses acting between adjacent pairs of beads are not correlated, since we have made the assumption that pair interactions are sufficient to describe the behaviour of the system.

\noindent
Under this assumption, we obtain
\begin{align*}
 \cexpval{\fprojbf_I \otimes \fprojbf_{I+1}}{\cvec{R}} &= \cexpval{ (\Delta\sigma_{I+1,I}-\Delta\sigma_{I,I-1}) \otimes (\Delta\sigma_{I+2,I+1}-\Delta\sigma_{I+1,I}) }{\cvec{R}} = \\
 &= -\cexpval{ \Delta\sigma_{I+1,I} \otimes \Delta\sigma_{I+1,I} }{\cvec{R}} \,, \\
 \\
 \cexpval{\fprojbf_I \otimes \fprojbf_{I-1}}{\cvec{R}} &= \cexpval{ (\Delta\sigma_{I+1,I}-\Delta\sigma_{I,I-1}) \otimes (\Delta\sigma_{I,I-1}-\Delta\sigma_{I-1,I-2}) }{\cvec{R}} = \\
 &= -\cexpval{ \Delta\sigma_{I,I-1} \otimes \Delta\sigma_{I,I-1} }{\cvec{R}} \,.
\end{align*}
Therefore we compute a smooth approximation of the variance of the fluctuating force between beads which allows us to define 
\begin{equation}
\label{eq:chi}
 \chi_{I+1,I} = \frac{\tau}{\tau_P}\,\frac{\Gamma\big(\tfrac{1}{\alpha}\big)}{\alpha}\,\beta\,\cexpval{ \Delta\sigma_{I+1,I} \otimes \Delta\sigma_{I+1,I} }{ \cvec{R} } \,,
\end{equation}
and the spatialy-varying friction coefficient matrix $\bs\chi$ (see \cref{Eq:approxM}) as
\begin{equation*}
\chi_{IJ}(\cvec{R}; \alpha, \tau) = \begin{cases}
    -\chi_{I+1,I} & I=J-1 \mod \ncg \,, \\
    \chi_{I+1,I} + \chi_{I,I-1} & I=J \,, \\
    -\chi_{I,I-1} & I=J+1 \mod \ncg \,,
 \end{cases}
\end{equation*}
where $I,J \in \bp{1,2,\dots,\ncg}$. We note here that the memory term we consider depends on CG coordinates. The importance of this feature in this kind of models was shown in \citep{Hudson2020}.

\subsection{Fitting and results}\label{Sec:Res}

In this section we show the numerical results for the autocorrelation function (ACF) of the fluctuating forces and of the bead momenta.

\begin{figure}[b]
\centering
 \includegraphics[width=0.45\textwidth]{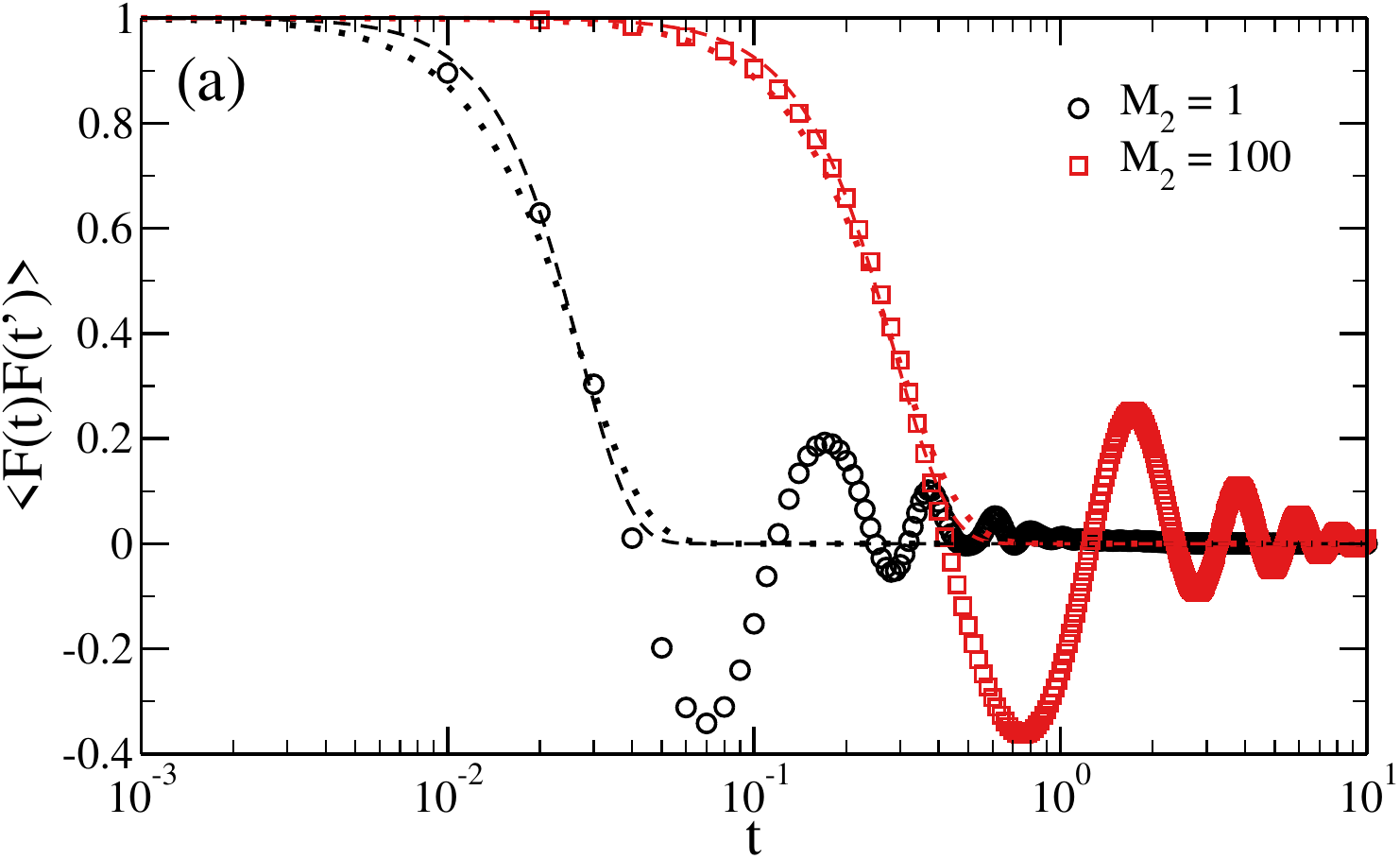}
 \includegraphics[width=0.45\textwidth]{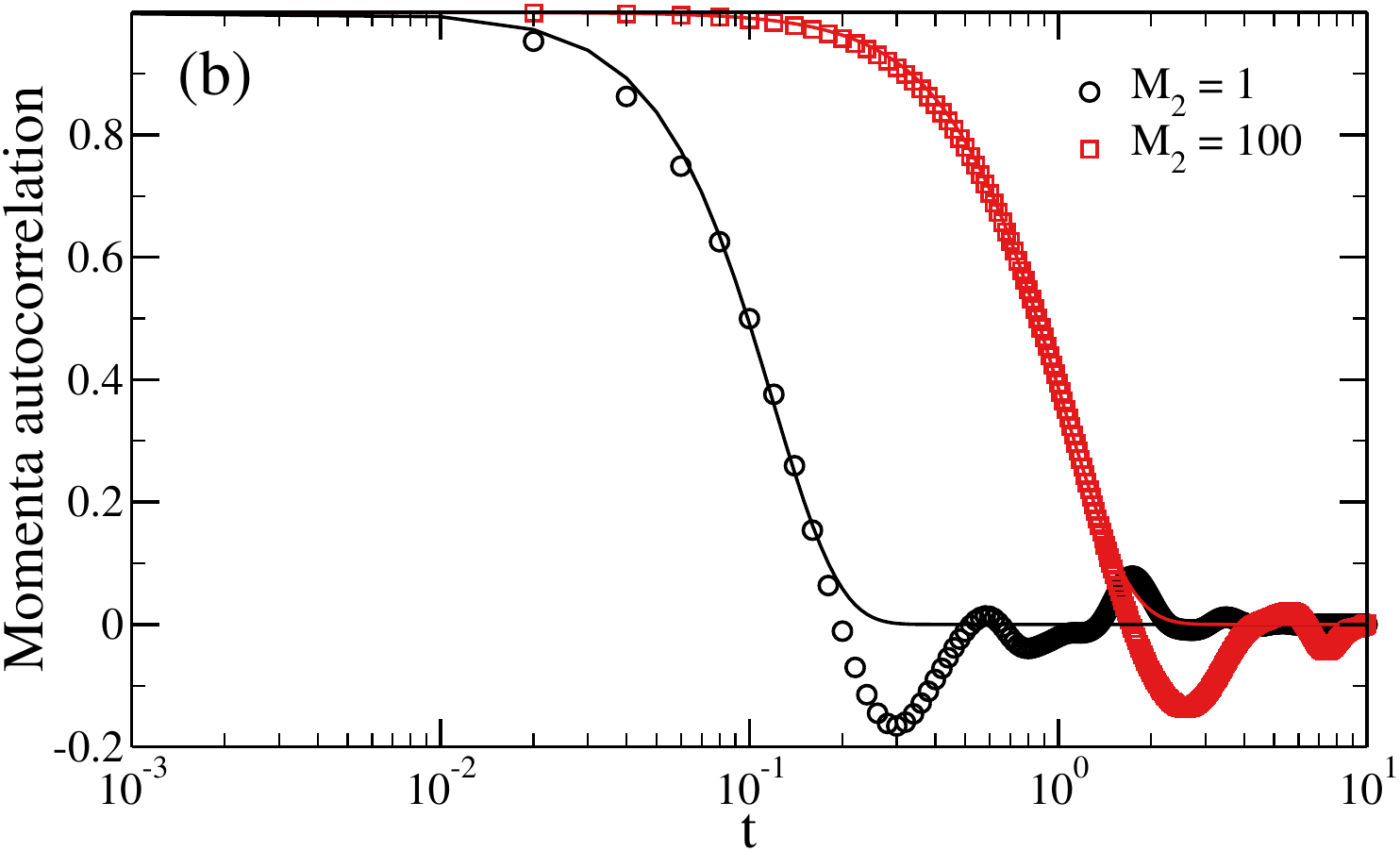}
 \caption{Time autocorrelation function of the (a) fluctuating force and (b) momenta for two values of $M_2$. Symbols represents the simulations data whereas dotted lines and dashed lines are fits to Eq.~\ref{eq:gaussian_fit} and Eq.~\ref{eq:general_fit}, respectively. In the ACF of the momenta (b) only one fitting is reported (solid black line) because Eq.~\ref{eq:gaussian_fit} and Eq.~\ref{eq:general_fit}  coincide.}
 \label{Fig:ACFcomparison}
\end{figure}

Applying our ansatz (see \cref{Eq:ansatz}) we fit the decaying of the ACF of the fluctuating forces obtained from a fine-grained simulation using the function
\begin{equation}
\label{eq:general_fit}
    F_F(t;\alpha_F,\tau) = \mbox{exp}\left[ -\cip{\frac{t}{\tau}}^{\alpha_F} \right].
\end{equation}
The best fit values for $\alpha_F$ and $\tau$, obtained with the XMGRACE software~\cite{turner2015xmgrace}, are reported in Table~\ref{tbl:fit_results} for the two systems investigated here. In order to check the consistency of the results, we also fit the initial decay of the ACF of the momenta and of the fluctuating forces using
\begin{equation} 
\label{eq:gaussian_fit}
    F_P(t;\tau_P) = \mbox{exp}\left[ -\cip{\frac{t}{\tau_P}}^2 \right]\,.
\end{equation}
For comparison, the data and the accompanying fitted curves are shown in \cref{Fig:ACFcomparison}.

\begin{table}[t]
\centering
\begin{tabular}{ |c|c|c|c| } 
 \hline
 $M_2$ & $\tau$ & $\alpha_F$ & $\tau_{P}$\\ 
 \hline
 1 & 0.027 & 2.59 & 0.12\\ 
 100 & 0.29 & 2.34 & 1.0\\ 
 \hline
\end{tabular}
\caption{Results of fitting to the numerical data.}
\label{tbl:fit_results}
\end{table}

For the fluctuating force, for both systems ($M_2 = 1$ and $M_2 = 100$) the initial behaviour can be described by a quasi-Gaussian behaviour with a characteristic decay time that does not depend on the exponent used, followed by damped oscillations that we do not take into account in our analysis. By contrast, the bead momenta ACF of both systems exhibit a nearly-perfect Gaussian decay. This distinction is likely due to the quadratic nature of the kinetic energy term in the Hamiltonian.

\begin{figure}[htb!]
\centering
 \includegraphics[width=0.8\textwidth]{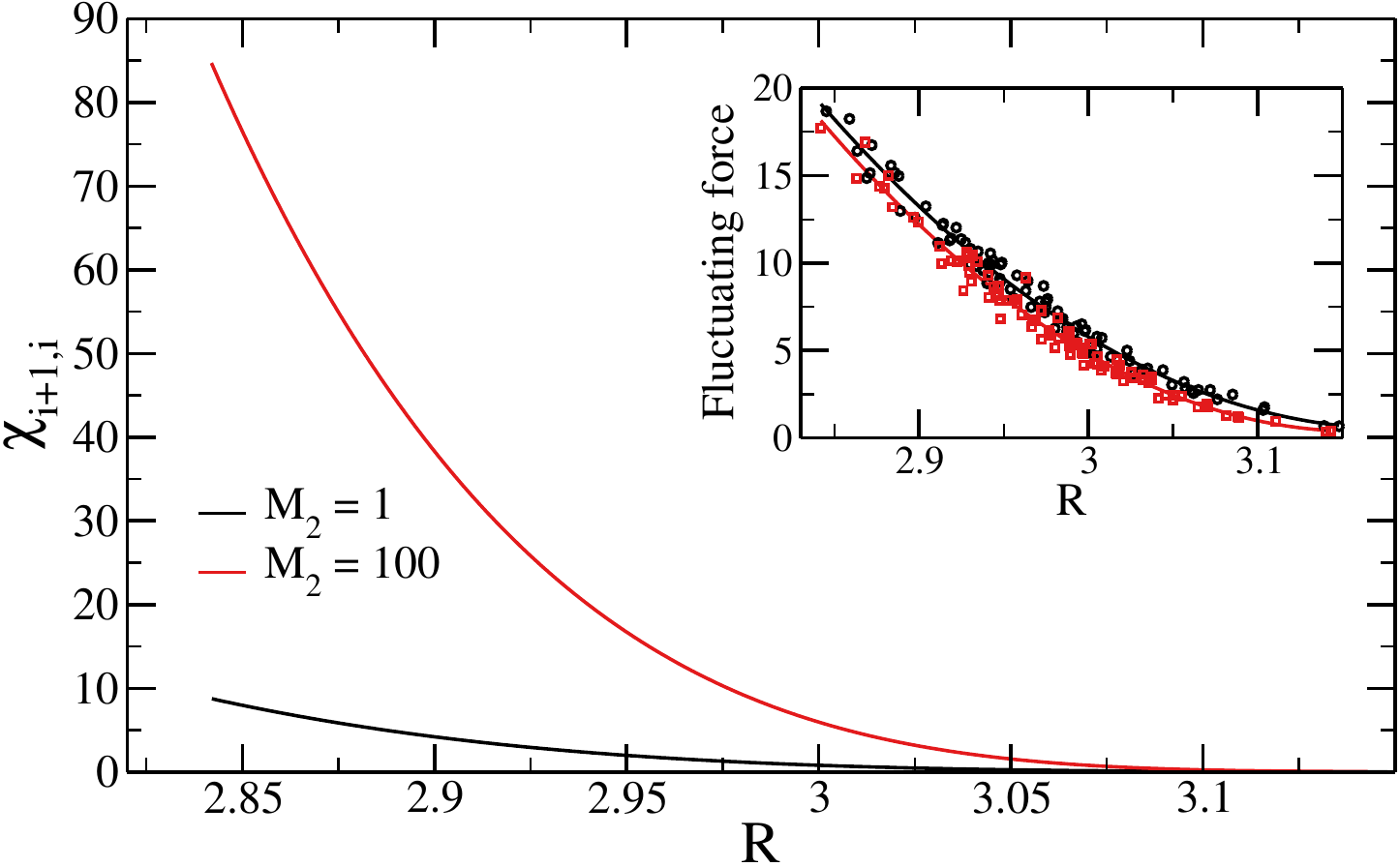}
 \caption{The friction coefficients evaluated for two values of $M_2$. The inset reports the fluctuating force for the same systems. The curves are spline interpolations of the numerical data, which is shown with symbols in the inset.}
 \label{Fig:FF_friction}
\end{figure}

We can now use \cref{eq:chi} to obtain the elements of $\bs\chi$, which explicitly depends on the inter-bead distance $\cvec{R}$ and are used to build our approximation of the memory term \cref{Eq:approxM}.
The results for the two systems investigated here are reported in Fig.~\ref{Fig:FF_friction}. The numerical data shows that, while the fluctuating forces $\fprojbf(0,\cdot)$ (shown in the inset of Fig.~\ref{Fig:FF_friction}) depends only weakly on the atomic masses, the resulting friction term features a much stronger dependence on $M_2$ through the value of $\tau$. This formulation demonstrates the possibility of using the result of Proposition~\ref{prop} to provide a CG approximation of the full system with appropriate theoretical guarantees on the error committed in replacing the memory integral with a simpler friction matrix.

\subsection{Discussion}

In the formulation we derived, we gave precise meaning to the concept of the scale separation and to the error committed in replacing the full memory term with its approximated value based on the ansatz reported in \cref{Eq:ansatz}.
In particular, this error is controlled by the two terms $\psi(t,\lambda)$ and $\zeta(\lambda)$ (see \cref{eq:psi,eq:zeta}). We also derived their asymptotic behaviour showing that  \cref{Eq:psiO,Eq:zetaO} are valid. In \cref{eq:relative} we derived the relative error of the approximation showing its boundary. It follows that, using the results obtained for \cref{Eq:psiO,Eq:zetaO}, we can write for the two terms of the boundary of the relative error in \cref{eq:relative}:
\begin{align}
    \zeta^\prime(\lambda) & = \frac{C_1}{\lambda^{\frac1\alpha}|G_0(\tilde{t})|}\frac{\Gamma\big(\frac2\alpha\big)}{\Gamma(\frac1\alpha)}  = \mathcal{O}\cip{\lambda^{-\frac{1}{\alpha}}} =  \mathcal{O}\cip{\frac{\tau}{\tau_P}} \label{Eq:Oz}\,, \quad \lambda\to\infty\,; \\
   \psi^\prime(\tilde{t},\lambda) & = \frac{\int_{\tilde{t}}^\infty \exp{-\lambda \tilde{s}^\alpha}\diff{\tilde{s}}}{\int_0^\infty \exp{-\lambda \tilde{s}^\alpha} \diff{\tilde{s}}}  = \mathcal{O}\cip{\exp{-\lambda \tilde{t}^{\alpha}}} =   \mathcal{O}\cip{\expp{-\cip{\frac{t}{\tau}}^{\alpha}}}  \label{Eq:Op} \,, \quad \tilde{t}\to\infty\,.  
\end{align}
%
%
We can now calculate the magnitude of the leading terms in \cref{Eq:Oz,Eq:Op} by using the quantities derived in preceding sections (see \cref{tbl:fit_results}).

In order to check the magnitude of the leading term in \cref{Eq:Op}, we need to define what is the time $t$. In the derivation presented in \cref{Sec:alphaPos}, $t$ represents the time scale of a CG simulation. For this reason, it seems reasonable to assume, for the sake of this calculation, that the typical time scale of the bead dynamics is represented by $\tau_P$.
Using the numbers reported in \cref{tbl:fit_results}, the leading term in \cref{Eq:Op} is $\expp{-\cip{\frac{\tau_P}{\tau}}^\alpha} \approx 10^{-21}$ for the system with $M_2 = 1$, and $\expp{-\cip{\frac{\tau_P}{\tau}}^\alpha} \approx 10^{-8}$ for the system with $M_2 = 100$. Although the specific values of these numbers (and even their orders of magnitude) depend on the functional form used to fit the simulation data, they are in all cases always much smaller than unity, demonstrating that the portion of the relative error bounded by $\psi^\prime(t)$ is extremely small. The term $\psi^\prime(t)$ is roughly related to the replacement of the integration boundary in \cref{eq:adim} from $(0,t)$ to $(0,\infty)$. It is therefore expected that the error due to this replacement decreases with the increase of $t$ and, given the argument of the integral, it decreases with exponential behaviour.

Let us now consider the portion of the relative error bounded by $\zeta^\prime(\lambda)$. With the same values reported in \cref{tbl:fit_results} we obtain $\frac{\tau}{\tau_P} = 0.225$ for the system with $M_2 = 1$ and $\frac{\tau}{\tau_P} = 0.29$ for the system with $M_2 = 100$. We see that, in this case, the bounding value for the percentage error is $\approx$ 23\% for the system with $M_2 = 1$  and  $\approx$ 30\% for the system with  $M_2 = 100$. The term $\zeta^\prime(\lambda)$ represents the relative error committed by replacing the memory term by its Taylor expansion stopped at a some $k$. In this case, we considered $k=0$, \textit{i.e.} we replaced the whole memory term with the leading term of its Taylor expansion. The value of the relative error bound in this case means that the results of the simulations should be carefully checked as it is not guaranteed that the error committed would be negligible. 

Since in \cref{Eq:zetaO} $\zeta(\lambda)$ approaches zero faster upon the increasing of $k$, this error bound can be reduced, making the results of such simulations more robust.

The analysis just showed puts on a more rigorous ground the qualitative statement that the characteristic time of the decay of the momenta ACF, being of the same order of magnitude of the decay time of the fluctuating-force ACF, makes resorting to \cref{eq:chi} questionable.
Moreover, our results show that including more terms in the approximation (see \cref{Sec:alphaPos}) should improve such model. We plan to test this approach in future work.

\section{Conclusions}

In this work we discussed the properties and various approximations of the dissipative memory term in the Mori-Zwanzig equations applied to Coarse-Grained simulations in molecular dynamics. 
A common approach to approximate this term is to assume a-priori a particular behaviour (\textit{e.g.}, Markovian) for the dynamics of the system and therefore simplify the terms involved in the memory accordingly. In this work, instead, starting from the Green-Kubo definition of the memory term involving the autocovariance of the fluctuating forces, we assume a functional form for this latter quantity. This is chosen such that an explicit calculation of the memory is possible, and it can be therefore compared and calibrated with Fine-Grained data. 

Starting from this ansatz, we proved a rigorous results on the boundary of the error committed by replacing the full memory term with its approximation. We also included an asymptotic analysis on the terms bounding the error and an explicit definition of the parameters which control the approximations used in the derivation. 

Finally, we tested this approach to a simple system represented by a one dimensional chain interacting with a Lennard-Jones type potential in two different cases, with different masses of the beads. We explicitly calculated the friction term for both of them checking the consistency of the Markovian approximation, using the asymptotic analysis presented in \cref{Sec:alphaPos}.

Our results show that this approach is feasible in a one-dimensional case, and demonstrates that the friction exhibits strong spatial dependence, suggesting that further investigation is required for more complex, multidimensional systems.

Moreover, our rigorous analysis clearly gives what is the range of error expected, showing that even for such a simple system like the one we considered the Markovian approximation should be checked carefully, as the error committed by such approximation can be important.

This work represents a step towards a systematic data-driven analysis of the memory term, and future work will focus on the analysis of the long-time memory behaviour of the system and the application to more realistic models.

\begin{appendices}

\crefalias{section}{appendix}

\setcounter{equation}{0}
\renewcommand{\theequation}{\thesection.\arabic{equation}}

\section{Proof of Proposition~\ref{prop}}\label{proof}
We begin by using Taylor's theorem (with remainder in Lagrange form) to expand $g$ in $s$ about $s=0$ up to oder $k-1$, writing
\begin{align*}
  g(t,s) &= \sum_{n=0}^{k} \frac{1}{n!}\frac{\partial^n}{\partial s^n}g(t,0)s^n+\frac{1}{k!}\bigg(\frac{\partial^k}{\partial s^k} g(t,\theta s)-\frac{\partial^k}{\partial s^k} g(t,0)\bigg)s^k\\
  &= \sum_{n=0}^{k} \frac{1}{n!}G_n(t)s^n+\frac{1}{k!}\bigg(\frac{\partial^k}{\partial s^k} g(t,\theta s)-\frac{\partial^k}{\partial s^k} g(t,0)\bigg)s^k
\end{align*}  
for some $\theta\in[0,1]$. Using this expansion and splitting the domain of integration, we have
\begin{multline*}
    \int_0^t \exp{- \lambda s^\alpha} g(t,s)\diff{s} - \sum_{n=0}^{k} \frac{G_n(t)}{n!}\int_0^{\infty} \exp{-\lambda s^\alpha} s^n\diff{s}\\
   = 
   \int_0^t \exp{- \lambda s^\alpha} \frac{1}{k!}\bigg(\frac{\partial^k}{\partial s^k} g(t,\theta s)-\frac{\partial^k}{\partial s^k} g(t,0)\bigg)s^k\diff{s} - \sum_{n=0}^{k} \frac{G_n(t)}{n!}\int_t^\infty \exp{-\lambda s^\alpha} s^n\diff{s}.
\end{multline*}
Employing the triangle inequality, the regularity assumption on $g$, and the fact that $\theta\in[0,1]$, we obtain
\begin{align*}
&\bigg|\int_0^t \exp{- \lambda s^\alpha} g(t,s)\diff{s} - \sum_{n=0}^{k} G_n(t)\int_0^{\infty} \exp{-\lambda s^\alpha} s^n\diff{s}\bigg|\\
   &\qquad\leq 
   \int_0^t \exp{- \lambda s^\alpha} \frac{1}{k!}\bigg|\frac{\partial^k}{\partial s^k} g(t,\theta s)-\frac{\partial^k}{\partial s^k} g(t,0)\bigg|s^k\diff{s} + \bigg|\sum_{n=0}^{k} \frac{G_n(t)}{n!}\int_t^\infty \exp{-\lambda s^\alpha} s^n\diff{s} \bigg|\\
   &\qquad\leq \frac{G_{k,\beta}(t)}{k!}\int_0^t \exp{- \lambda s^\alpha} s^{k+\beta}\diff{s} + \bigg|\sum_{n=0}^{k} \frac{G_n(t)}{n!}\int_t^\infty \exp{-\lambda s^\alpha} s^n\diff{s}\bigg|\\
   &\qquad\leq \frac{G_{k,\beta}(t)}{k!}\int_0^\infty \exp{- \lambda s^\alpha} s^{k+\beta}\diff{s} + \sum_{n=0}^{k} \frac{\big|G_n(t)\big|}{n!}\int_t^\infty \exp{-\lambda s^\alpha} s^n\diff{s}\,,
\end{align*}
where $G_{k,\beta}(t)$ is the H\"older $\beta$-semi-norm of a bounded function $g$ on the set $[0,t]$.
We note that, for the evolution over a given time interval $[0,T]$, we can replace the functions $G_n(t)$ and $G_{k,\beta}(t)$ by appropriate constants $C_n$ and $C_{k+1}$ by taking a supremum over the time interval.

To conclude, we note that by making the change of variable, we can express the integrals over the interval $(0,\infty)$ in terms of the Gamma function:
\[
\int_0^\infty \exp{-\lambda s^\alpha} s^\gamma \diff{s} = \frac{\Gamma\Big(\frac{\gamma+1}{\alpha}\Big)}{\alpha \lambda^{\frac{\gamma+1}{\alpha}}}.
\]
Using this result, we conclude that the statement holds.

\section{Proof of \cref{eq:psi}}\label{app:psi}

Let us start with the definition of $\psi(t)$

\begin{align}
    \psi(t) & \overset{\operatorname{def}}{=} \sum_{n=0}^{k}\frac{C_n}{n!}\int_t^\infty \exp{-\lambda s^\alpha} s^n \diff{s} \nonumber \\
    & \leq (k+1) \max_{n \in \{0,1,\ldots,k\}}\sqp{ \frac{C_n}{n!}\int_t^\infty \exp{-\lambda s^\alpha} s^n \diff{s}}.
\end{align}

Let us now estimate the last integral in the previous expression. 
We start by making the change of variable $v = s^\alpha$:
\begin{align*}
  \int_t^\infty \exp{-\lambda s^\alpha} s^n\diff{s}
  = \int_{t^\alpha}^\infty \frac{1}{\alpha} \exp{-\lambda v} v^{\frac{n+1}{\alpha}-1}\diff{v}.
\end{align*}
There are now two cases: either (i) $\frac{n+1}{\alpha}-1\leq 0$, or (ii) $\frac{n+1}{\alpha}-1> 0$.\\
In case (i), we can estimate the integral as being
\begin{gather*}
  \int_t^\infty \exp{-\lambda s^\alpha} s^n\diff{s}
  = \int_{t^\alpha}^\infty \frac{1}{\alpha} \exp{-\lambda v} v^{\frac{n+1}{\alpha}-1}\diff{v}
  \leq
  \\ 
 \qquad \frac{t^{n+1-\alpha}}{\alpha}\int_{t^\alpha}^\infty  \exp{-\lambda v} \diff{v}=
  \frac{t^{n+1}}{\alpha}\frac{\exp{-\lambda t^{\alpha}}}{\lambda t^{\alpha}} \,.
\end{gather*}
This is exponentially small as long as $\lambda t^\alpha\gg 1$. 

In case (ii), we can integrate by parts to obtain
\[
  \int_{t^\alpha}^\infty \frac{1}{\alpha}\,\exp{-\lambda v}\,v^{\frac{n+1}{\alpha}-1}\,\diff{v} 
   = -\frac{1}{\alpha} \left( \frac{ \exp{-\lambda t^\alpha} }{ \lambda t^\alpha } \,t^{n+1}
  - \cip{\frac{ \tfrac{n+1}{\alpha}-1 }{\lambda}}\int_{t^\alpha}^\infty \,\exp{-\lambda v} \,v^{\frac{n+1}{\alpha}-2} \,\diff{v}\right).
\]
Integrating by parts more times as necessary, we ultimately obtain a similar result to that obtained in case (i), \textit{i.e.} that this error term is exponentially small as long as $\lambda t^\alpha\gg 1$.

In summary, we have shown that 
\begin{equation}\label{Eq:psiO}
   \psi(t)  = \mathcal{O}\cip{\exp{-\lambda t^{\alpha}}}\,, \quad t\to\infty.  
\end{equation}

\end{appendices}


\bibliographystyle{chicago}
\bibliography{bibliography_20Feb20}
\end{document}